\documentclass[final,3p,times,twocolumn]{elsarticle}

\usepackage{amssymb,amsmath,amsthm,bm}
\usepackage{geometry}
\usepackage{algorithm}
\usepackage{algpseudocode}

\usepackage{graphicx}
\usepackage[retainorgcmds]{IEEEtrantools}
\graphicspath{{figures-pdf/}}
\usepackage[normalem]{ulem}
\usepackage{url}
\usepackage{lineno}

\usepackage{cases}
\usepackage{slashbox}

\usepackage{bm}
\usepackage{graphicx,color,overpic}

\usepackage{float}

\newlength\figsep
\let\labelindent\relax    
\usepackage{enumitem}
\newlength\OneImW
\newlength\TwoImW
\newlength\BigOneImW
\newlength\ThreeImW
\newlength\DoubleThreeImW
\newlength\sfigwidth
\newlength\vfigskip
\journal{IEEE Multimedia}
\usepackage{color}
\definecolor{dgreen}{rgb}{0,.6,0}

\newtheorem{Proposition}{Proposition}

\usepackage[font=small,skip=0pt]{caption}
\emergencystretch=\maxdimen
\usepackage{hyperref}
\hypersetup{
 linktocpage=true,
 pdfborderstyle={/S/S/W 1},
 hyperindex=true,
 bookmarks=true,
 bookmarksopen=true,
 bookmarksnumbered=true,
}

\begin{document}

\begin{frontmatter}


\title{Cryptanalyzing an image encryption algorithm based on autoblocking and electrocardiography}

\author[hnu-cn]{Chengqing Li\corref{corr}}
\ead{DrChengqingLi@gmail.com}

\author[hnu-cn]{Dongdong Lin}

\author[cas-cn]{Jinhu L\"u}

\author[ncl-uk]{Feng Hao}

\cortext[corr]{Corresponding author.}

\address[hnu-cn]{School of Computer Science and Electronic Engineering, Hunan University, Changsha 410082, Hunan, China}

\address[cas-cn]{Academy of Mathematics and Systems Sciences, Chinese Academy of Sciences, Beijing 100190, China}


\address[ncl-uk]{School of Computing Science, Newcastle University, Newcastle Upon Tyne NE1 7RU, UK}

\begin{abstract}
This paper analyzes the security of an image encryption algorithm proposed by Ye and Huang [\textit{IEEE MultiMedia}, vol. 23, pp. 64-71, 2016]. The Ye-Huang algorithm uses electrocardiography (ECG) signals to generate the initial key for a chaotic system and applies an autoblocking method to divide a plain image into blocks of certain sizes suitable for subsequent encryption. The designers claimed that the proposed algorithm is ``strong and flexible enough for practical applications''. In this paper, we perform a thorough analysis of their algorithm from the view point of modern cryptography. We find it is vulnerable to the known plaintext attack: based on one pair of a known plain-image and its corresponding cipher-image, an adversary is able to derive a mask image, which can be used as an equivalent secret key to successfully decrypt other cipher-images encrypted under the same key with a non-negligible probability of 1/256. Using this as a typical counterexample, we summarize security defects in the design of the Ye-Huang algorithm. The lessons are generally applicable to many other image encryption schemes.

\end{abstract}
\end{frontmatter}

\section{Introduction}

Security and privacy of image data have become almost everyone's concern as sharing and enjoying photos on
social media are a part of our daily lives nowadays, which is strongly supported by human's complex emotional needs, e.g., narcissism, popularity and belongingness. To cope with the challenges, a great number of image encryption and privacy protection schemes
were proposed to conceal important information about the original image data from the unintended viewers \cite{zhouyc:share:TIP17,Cqli:Scramble:IM17,Cqli:Fridrich:SP2017}. The complex dynamics of chaotic maps demonstrated in an infinite-precision world are similar to the required properties of a secure encryption system initially summarized by Shannon, the father of information theory,
in the late 1940s \cite[Table 1]{LiShujun:Rules:IJBC2006}. The similarity attracts many security researchers to utilize various chaotic systems and methodologies for all kinds of cryptographic applications, including image encryption, video encryption, image privacy protection, public key infrastructure and hash. Some biometric personal features, e.g. fingerprint, iris, and electrocardiography (ECG) signals, are used for identification and cryptography 
in various application scenarios, e.g. Internet of Things \cite{peris2018effect:ECG}.

ECG records electrical changes of the skin arising from the heart muscle's electrophysiologic patterns of depolarization and repolarization during each heartbeat. The dynamic properties of the time series of ECG, measured by the 2D degree distribution of the corresponding complex networks, can be used as a tool to identify healthy persons from pathological groups. In different cryptographic scenarios, the ECG signal is used for various purposes. In \cite{Zhang:ECG:ITITB2012}, its characteristics in Fourier domain
are used as a key for realizing secure communication and hash-based authentication among sensor nodes in a body area sensor networks (BANs).
To assure an ECG signal is securely transmitted in BANs, it is first compressed with a compression algorithm called SPIHT (set partitioning in hierarchical trees) and then a small portion of important coefficients in the compression domain are encrypted using a well-known modern cipher \cite{Ma:ECG:ITBE2012}. The importance of the compression coefficients is evaluated in terms of their influence strength on decompression. To keep a patient's ECG information private in an automatic online diagnosis system identifying six possible states of heart beat, four feature coefficients of the ECG signal are extracted and encrypted by a homomorphic encryption scheme before transmission to the system \cite{Barni:ECG:ITIFS2011}. In \cite{Safie:ECG:TIFS2011}, a feature extraction technique of ECG is proposed to accurately authenticate whether two ECG signals belong to the same person.

In \cite{chen2012personalized}, a personalized information encryption scheme using ECG signals with chaotic functions was proposed. In this scheme, the
Lyapunov exponent of an ECG signal is used as the initial states of two pseudorandom number generators composed of a logistic map and a Henon map, respectively.
As selection of chaotic map is a core step in the design of a chaos-based cryptosystem, some unimodal maps like logistic map can weaken security of the supported cryptosystem \cite{Arroyo:logistic:2008}. In addition, dynamics of any chaotic map in the digital domain are degenerated and may seriously influence the security of the supporting encryption function \cite{alvarez2011lessons,WangQX:HDDCS:TCAS2016}. The paper \cite{chen2012personalized} skipped the problem and did not provide any information on the concrete functions used for diffusion and confusion, which violates some basic design principles summarized in \cite{LiShujun:Rules:IJBC2006}. To improve the scheme proposed in \cite{chen2012personalized}, Ye and Huang proposed an image encryption algorithm based on autoblocking and electrocardiography (IEAE) in \cite{ye2016image}. Their method is to use an ECG signal to generate the initial keys to control the whole encryption process composed of block-wise matrix multiplications.
This paper re-evaluates the security of IEAE and we find that IEAE is susceptible to a known-plaintext attack. In addition, the security defects in IEAE are summarized, along with lessons for avoiding similar pitfalls in the design of image encryption schemes.

The rest of the paper is organized as follows. Section~\ref{scheme} presents a description of IEAE.
Detailed cryptanalytic results on IEAE are given in Section~\ref{cryptanalysis}. The last section concludes the paper.

\section{Image encryption algorithm based on autoblocking and electrocardiography (IEAE)}
\label{scheme}

The encryption object of IEAE is a gray-scale image of size $M\times N$, denoted by $\mathbf{I}=\{ I_{i, j} \}_{i=1, j=1}^{M, N}$. The whole plain-image is divided into blocks of size $p_1 \times p_2$ and is encrypted blockwise by IEAE.
Let $\mathbf{C}$ denotes the corresponding cipher-image. Then, the basic parts of IEAE can be described as follows.

\begin{itemize}[leftmargin=*]

\item \textit{The secret key}: non-negative integers $\omega_1$, $\omega_2$, $\mu_1$, $\mu_2$ used for array indexes; control parameter of Logistic map
\begin{equation}
\label{eq:LogisticMap}
x_n=\mu \cdot x_{n-1}\cdot (1-x_{n-1}),
\end{equation}
$\mu\in[3.9, 4]$; positive integer control parameters of generalized Arnold map
\begin{equation}
\begin{pmatrix}
  x_{n+1} \\
  y_{n+1}
\end{pmatrix}
=
\begin{pmatrix}
    1 & a     \\
    b & 1+a\cdot b
\end{pmatrix}\cdot
\begin{pmatrix}
  x_{n} \\
  y_{n}
\end{pmatrix}
\bmod 1,
\label{eq:Arnold}
\end{equation}
$a$ and $b$,
where $(x \bmod n)=x- n \cdot \lfloor x/n \rfloor$.

\item \textit{Public parameter}: iteration number $R$.

\item \textit{Initialization}:

1) Given an ECG signal $\mathbf{Z}=\{ z_i \}_{i=1}^L$, its largest Lyapunov exponent, $\lambda$,
is calculated by Wolf's algorithm proposed in \cite{wolf1985determining}:
\begin{itemize}[leftmargin=*]
  \item \textit{Step 1}: Transform the signal $\mathbf{Z}$ into a sequence in an $m$-dimensional phase space, $\mathbf{Y}=\{ Y_i \}_{i=1}^{L^*}$,
  where
  \begin{equation*}
  Y_i =[ z_{(i-1)m+1}, z_{(i-1)m +2}, \cdots, z_{(i-1)m+m} ],
  \end{equation*}
  $L^*=\lfloor L/m \rfloor$. Initialize indexes $k=1$ and $t_k=1$.

 \item \textit{Step 2}: Calculate the distance
 \begin{equation*}
 L_k=\| Y_{t_k}-Y_{t'_k}\|,
 \end{equation*}
 where $Y_{t'_k}$ is the directional nearest neighbor point of $Y_{t_k}$ in the phase space, and the angle between $\overrightarrow{Y_{t_k}Y_{t'_{k-1}}}$ and $\overrightarrow{Y_{t_k}Y_{t'_k}}$, $\theta$, is smaller than $30^{\circ}$ when $k>1$.

 \item \textit{Step 3}: As the evolution and replacement procedure depicted in Fig.~\ref{fig:wolfLyapunov}, incrementally increase the values of $t_k$ and $t'_k$ at the same time until the distance between
 $Y_{t_k}$ and $Y_{t'_k}$ is larger than the threshold value $\epsilon$. Then, set their current distance as $L'_k$, $k=k+1$,
 and $t_k=t_{k-1}$.

   \item \textit{Step 4:} Repeat \textit{Step 2} and \textit{Step 3} until $t_k>L^*$.

  \item \textit{Step 5:} Calculate the largest Lyapunov exponent
  \begin{equation}
    \lambda = \frac{1}{t_k-1}\sum_{k=1}^q \log_2\left( \frac{L'_k}{L_k} \right),
    \label{lyapunov}
  \end{equation}
where $q$ is the total number of the replacement steps.
\end{itemize}

\begin{figure}[!htb]
\centering
\begin{minipage}{\BigOneImW}
\centering
\includegraphics[width=\BigOneImW]{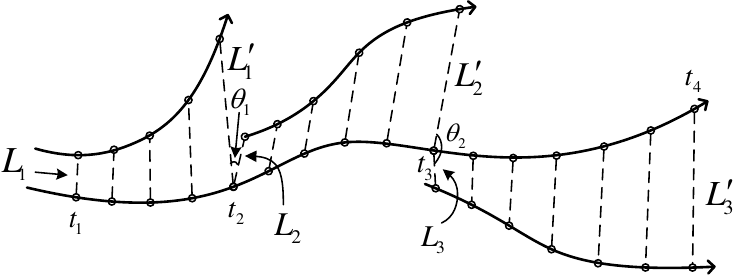}
\end{minipage}
\caption{Schematic representation of the evolution and replacement procedure estimating the largest Lyapunov exponent from a phase space.}
\label{fig:wolfLyapunov}
\end{figure}

2) Iterate Eq.~\eqref{eq:LogisticMap} $P$ steps from initial condition $\bar{x}_0= \mathrm{Rem}(|\lambda| \cdot 10^8)$ with the control parameter $\mu$ and obtain integer sequence $\{ \bar{x}_1, \bar{x}_2, \cdots, \bar{x}_P \}$ via the conversion function
\begin{equation}
f(x)=\left( x \cdot 10^{14} \right) \bmod 256,
\label{eq:conversion}
\end{equation}
where $P=(p_1\cdot p_2)+256$, and $\mathrm{Rem}(x)$ returns the fractional part of $x$.

3) Iterate Eq.~\eqref{eq:Arnold} $Q$ steps from $(x_0, y_0)=(|\lambda|, \mathrm{Rem}(|\lambda| \cdot 10^5))$ and assign the obtained sequence $\{ x_{r+1}, y_{r+1}, x_{r+2}, y_{r+2},\cdots x_{r+MN/2}, y_{r+MN/2} \}$ converted by Eq.~(\ref{eq:conversion}) into $M \times N$ matrix $\mathbf{D}$ in the raster order, where $r=\bar{x}_{\mu_1}$, and $Q=r+MN/2$.

4) Assign sequence $\{ \bar{x}_{\mu_3}, \bar{x}_{\mu_3+1}, \cdots \bar{x}_{\mu_3+p_1\cdot p_2} \}$ into $p_1\times p_2$ matrix $\mathbf{C}_0$ in the raster order, where
\begin{equation}
\mu_3=\left( \sum_{\substack{i=1, j=1}}^{p_1, p_2}I_{i, j} \right)\mod 256+1,
\label{eq:linkplaintext}
\end{equation}

\item \textit{The encryption procedure}:

\begin{itemize}[leftmargin=*]

\item \textit{Step 1}: Divide $\mathbf{I}$ and $\mathbf{D}$ into $r_1\cdot r_2$ sub-blocks of size $p_1\times p_2$, which is automatically
selected from a fixed look-up table via random entries. Table~\ref{tb:blocksize} shows the table used for plain-images of size $256\times 256$.
In this case, indexes
\begin{equation}
\label{eq:multiply14}
\begin{cases}{}
 q_1=\lfloor \bar{x}_{\omega_1}\cdot 10^{14} \rfloor \bmod 3,\\
 q_2=\lfloor \bar{x}_{\omega_2}\cdot 10^{14} \rfloor \bmod 3.
\end{cases}
\end{equation}
For example, the block size is set as $(p_1, p_2)=(8, 16)$ if $(q_1, q_2)=(0, 1)$.
If $p_1\nmid M$ or $p_2\nmid N$, some zero pixels are padded to the plain-image to make
equations $M=r_1\cdot p_1$ and $N=r_2\cdot p_2$ both exist.

\begin{table}[!htb]
\centering
\caption{Block sizes for plain-image of size $256\times 256$.}
\label{tb:blocksize}
\begin{tabular}{c|ccc}  \hline
\backslashbox{ $q_1$ }{ $q_2$ }    & 0      & 1       & 2       \\ \hline
  0 & (8,8)  & (8,16)  & (8,32)  \\
  1 & (16,8) & (16,16) & (16,32) \\
  2 & (32,8) & (32,16) & (32,32) \\ \hline
\end{tabular}
\end{table}

\item \textit{Step 2}: Encrypt the $k$-th block of plain-image $\mathbf{I}$, $I_k$, by
\begin{equation}
 \label{eq:encrypt_ori}
 C_k =
 \begin{cases}
   (I_k + v\cdot D_k + C_{k-1}) \bmod 256 & \text{if } k < r_1\cdot r_2; \\
   (I_k + C_{k-1})        \bmod 256       & \text{if } k = r_1\cdot r_2,
 \end{cases}
\end{equation}
for $k=1 \sim (r_1\cdot r_2)$, where $v=\bar{x}_{\mu_2}$, and $D_k$ denotes the $k$-th block divided in Step 1).
To facilitate the following description, we unify the two functions in Eq.~(\ref{eq:encrypt_ori})
as the same form,
\begin{equation}
 \label{eq:encrypt}
C_k  = (I_k + v\cdot D_k + C_{k-1}) \bmod 256,
\end{equation}
by setting $D_k\equiv 0$ if $k=r_1\cdot r_2$.

\end{itemize}

\item \textit{Step 3}: Repeat the above step $R$ times.

\item \textit{The decryption procedure} is the inverse version of Eq.~(\ref{eq:encrypt}), and operates
\begin{equation}
I_k =
\begin{cases}
(C_k - v\cdot D_k - C_{k-1})\bmod 256 & \text{if } k < r_1\cdot r_2; \\
(C_k - C_{k-1}) \bmod 256             & \text{if } k = r_1\cdot r_2,
\end{cases}
\end{equation}
for $k=(r_1\cdot r_2)\sim 1$.

\end{itemize}

\section{Cryptanalysis of IEAE}
\label{cryptanalysis}

In \cite{ye2016image}, the authors claimed that ``the keystream generated is
related to the plain-image, so it can effectively resist all kinds of differential attacks."
However, we argue that the statement is not true. Furthermore, we report the underlying mechanisms relating to the insecurity of some
basic parts of IEAE.

\subsection{Known-plaintext attack on IEAE}

The known-plaintext attack is a cryptanalysis model assuming that the attacker can access a set of plaintexts and the corresponding ciphertexts
encrypted by the same secret key. If an encryption scheme cannot withstand known-plaintext attack, every secret key should only be used only once in one encryption session, which
would incur complex management of secret keys and very high cost. According to Kerckhoffs's principle, an encryption algorithm should be secure even if everything about the algorithm, except the secret key, is public knowledge, which is reformulated  by Shannon as ``the enemy knows the system". In \cite{ye2016image}, the authors claimed that ``known-plaintext and chosen-plaintext attacks are infeasible for the proposed
encryption algorithm" based on the sensitivity of matrix $\mathbf{C}_0$ on the change of plain-images, caused by the mechanism shown in Eq.~(\ref{eq:linkplaintext}). Actually, the sensitivity mechanism is cancelled due to the modulo addition in Eq.~(\ref{eq:linkplaintext}), which occurs with probability $1/256$ if every pixel in the plain-images is distributed uniformly.

\begin{Proposition}
Given any round number $R$,
$\mathbf{D}'=\{D'_k\}_{k=1}^{r_1r_2}$
is the equivalent secret key of IEAE for other plain-images generating the same value of $\mu_3$, where
\begin{equation}
  \label{eq:kpa}
  D'_k = \left( C_k - \underbrace{\sum_{h_1=1}^k \sum_{h_2=1}^{h_1} \cdots \sum_{h_{R-1}=1}^{h_{R-2}} \sum_{i=1}^{k-h_{R-1}+1}}_{\text{$R$ times}} I_i \right) \bmod 256.
\end{equation}
\label{prop:equivalentKey}
\end{Proposition}
\begin{proof}
Observing Eq.~(\ref{eq:encrypt}), one has
\begin{IEEEeqnarray*}{rCl}
 C_k &= & (I_k + v\cdot D_k + C_{k-1}) \bmod 256 \\
     &= & (I_k + I_{k-1} + v\cdot (D_k + D_{k-1}) + C_{k-2}) \bmod 256 \\
     & \vdots &  \\
     &= &\left( \sum_{i=1}^k I_i + v\cdot \sum_{i=1}^k D_i + C_0 \right) \bmod 256,
\end{IEEEeqnarray*}
for any $k \in \{1, 2, \cdots, r_1\cdot r_2 \}$.
When $R=2$ and the relationship between $C_k$ and $I_k$ becomes
\begin{IEEEeqnarray*}{rCl}
  C_k & = &  \left( \sum_{i=1}^k I_i + v\cdot \sum_{i=1}^k D_i + C_0 + v\cdot D_k + C_{k-1} \right) \bmod 256\\
      & = &  \left( \sum_{i=1}^k I_i + v\cdot \sum_{i=1}^k D_i + C_0 + v\cdot D_k +  \right.\\
      &   &  \left. \; \sum_{i=1}^{k-1} I_i + v\cdot \sum_{i=1}^{k-1} D_i + C_0 + v\cdot D_{k-1} + C_{k-2} \right) \bmod 256\\
      & = & \left( \sum_{h_1=1}^2\! \sum_{i=1}^{k-h_1+1}\! I_i + v\cdot\! \sum_{h_1=1}^2\! \sum_{i=1}^{k-h_1+1}\! D_i +   2 C_0 +   \right.\\
      &   & \left. \; v\cdot\! \sum_{i=k-1}^k\! D_i+  \cdots  + I_1+ v\cdot D_1 + C_0  \right) \bmod 256,\\
      & = & \left( \sum_{h_1=1}^k\! \sum_{i=1}^{k-h_1+1}\! I_i + v\cdot\! \sum_{h_1=1}^k\! \sum_{i=1}^{k-h_1+1}\! D_i + v\cdot\! \sum_{i=1}^k\! D_i + k\cdot C_0 \right)\! \bmod 256.
\end{IEEEeqnarray*}

As for any value of round number $R$, the relationship can be similarly derived:
\begin{equation}
\label{eq:encrypt_e}
C_k = \left( \underbrace{\sum_{h_1=1}^k \sum_{h_2=1}^{h_1} \cdots \sum_{h_{R-1}=1}^{h_{R-2}} \sum_{i=1}^{k-h_{R-1}+1}}_{\text{$R$ times}} I_i + D'_k \right) \bmod 256,
\end{equation}
where
\begin{IEEEeqnarray*}{rCl}
  \label{eq:encrypt_r_1}
 D'_k & = &  \left( v\cdot   \left(    \underbrace{\sum_{h_1=1}^k  \sum_{h_2=1}^{h_1}  \cdots  \sum_{h_{R-1}=1}^{h_{R-2}}  \sum_{i=1}^{k-h_{R-1}+1}}_{\text{$R$ times}} D_i   \right. \right.\\
      &  & +\left. \underbrace{\sum_{h_1=1}^k  \sum_{h_2=1}^{h_1}  \cdots  \sum_{h_{R-2}=1}^{h_{R-3}}  \sum_{i=1}^{k-h_{R-2}+1}}_{\text{$R-1$ times}} D_i
  + \cdots  + \sum_{i=1}^k D_i   \right)\\
      &  & + \left. \underbrace{\sum_{h_1=1}^k \sum_{h_2=1}^{h_1} \cdots \sum_{h_{R-3}=1}^{h_{R-4}} \sum_{i=1}^{k-h_{R-3}+1}}_{\text{$R-2$ times}} i\cdot C_0 \right) \bmod 256,
\end{IEEEeqnarray*}
From Eq.~(\ref{eq:encrypt_e}), one can see that the mask image $\mathbf{D}'=\{D'_k\}_{k=1}^{r_1r_2}$
can work as the equivalent key of IEAE, which completes the proof of this proposition.
\end{proof}

\begin{figure}[!htb]
\centering
\begin{minipage}[t]{\ThreeImW}
\centering
\includegraphics[width=\ThreeImW]{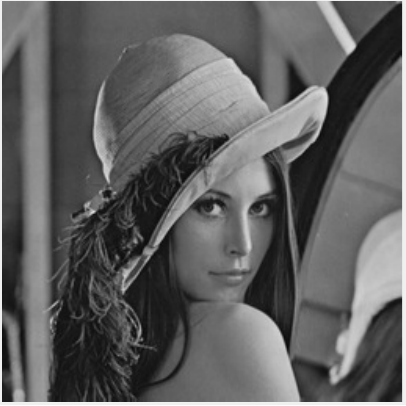}
a)
\end{minipage} \hspace{3pt}
\begin{minipage}[t]{\ThreeImW}
\centering
\includegraphics[width=\ThreeImW]{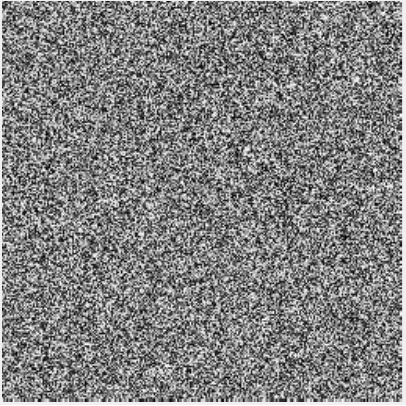}
b)
\end{minipage}
\hspace{3pt}
\begin{minipage}[t]{\ThreeImW}
\centering
\includegraphics[width=\ThreeImW]{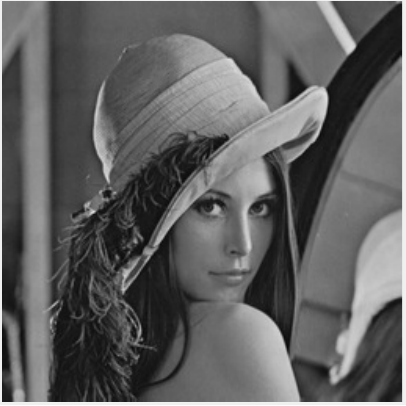}
c)
\end{minipage}\\
\begin{minipage}[t]{\ThreeImW}
\centering
\includegraphics[width=\ThreeImW]{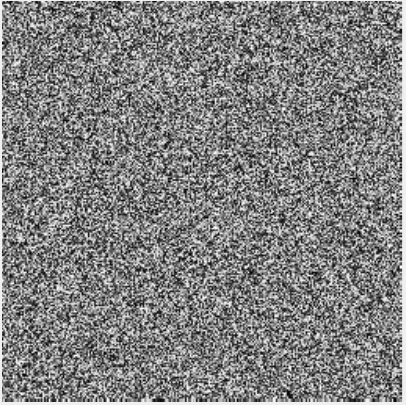}
d)
\end{minipage}
\hspace{3pt}
\begin{minipage}[t]{\ThreeImW}
\centering
\includegraphics[width=\ThreeImW]{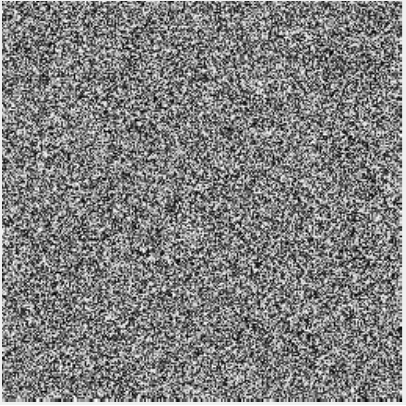}
e)
\end{minipage}
\hspace{3pt}
\begin{minipage}[t]{\ThreeImW}
\centering
\includegraphics[width=\ThreeImW]{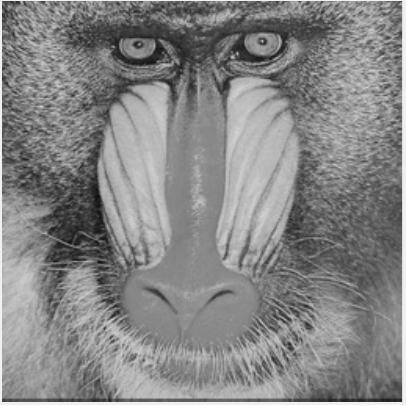}
f)
\end{minipage}
\caption{The results of IEAE cryptosystem and the proposed known-plaintext attack method: a) ``Lenna''; b) encrypted ``Lenna''; c) decrypted ``Lenna'';
d) encrypted ``Baboon'' with the same secret key; e) mask image $\mathbf{D}'$; f) the breaking result of the encrypted ``Baboon.''}
\label{fig:encryptAndAttack}
\end{figure}

Given one pair of plain-image and the corresponding cipher-image, a mask image can be constructed as Proposition~\ref{prop:equivalentKey}, which
can be used to decrypt a cipher-image when the following two conditions hold at the same time: 1) it is encrypted by IEAE with the same secret key and
public parameter as the given cipher-image; 2) its corresponding plain-image can generate the same value of $\mu_3$ as the given plain-image. To check the validity of the proposed attack, we performed a number of experiments with some secret keys and more than 256 plain-images of size $512\times 512$.
When the secret key and parameter are set as \cite{ye2016image}, namely $a=1$, $b=1$, $\omega_1=50$, $\omega_2=50$, $\mu=3.999$, $\mu_1=20$, $\mu_2=15$, and $R=3$, the plain-image ``Lenna" and its results of encryption and decryption are shown in
Fig.~\ref{fig:encryptAndAttack}a), b), and c), respectively. A number of cipher-images encrypted with the same secret key
as that of Fig.~\ref{fig:encryptAndAttack}b) were decrypted by using the equivalent secret key obtained by the above attack, which is
shown in Fig.~\ref{fig:encryptAndAttack}e). Among them, the cipher-image shown in Fig.~\ref{fig:encryptAndAttack}d) was successfully decrypted since the corresponding plain-image shown in
Fig.~\ref{fig:encryptAndAttack}f) can generate the same $\mu_3$ as the plain-image ``Lenna" by Eq.~(\ref{eq:linkplaintext}).

\subsection{Security defects of IEAE}

Using IEAE as a representative example, we analyze here the underlying mechanisms for its
security defects, which also exist in many other image encryption schemes.

\begin{itemize}[leftmargin=*]

\item The real structure of Logistic map in digital computer

In a finite-precision digital computer, dynamics of any chaotic maps satisfying a well-known chaos definition in the infinite domain
will definitely be degraded. Reliability of numerical solution of some chaotic dynamical systems is questionable \cite{lozi2013trustChaos}.
Given an arithmetical domain (all possible representable numbers) and a rounding method, the \textit{functional graph} of a chaotic map is determined. Just as \cite[Fig.~1]{ye2016image}, a great number of research papers use the change trend
of the positive Lyapunov exponent of digital chaotic maps with respect to control parameters to demonstrate complex degree of their dynamics.
The metric only measures the maps from the macroscopic perspective. Actually, the calculated Lyapunov exponent in the digital domain is the change trend of the underlying functional graph along some evolution orbits (paths). So, some subtle properties of the digital chaotic map are omitted \cite{cqli:networkDynamics17}, such as short period cycles. To show this point, Fig.~\ref{fig:networkLogistic6bitsfloorroundceil} depicts the functional graph of Logistic map with $\mu=61/2^4$ in the arithmetic domain $\{0, 1, 2, \cdots, 2^6\}$ under three quantization methods. Some small-sized connected components are omitted with a relatively high probability, which may lead to security defects of the supported system.
Note that the structure of the functional graph of Logistic map implemented in a high precision is very similar to that in a lower precision \cite{cqli:networkDynamics17}. Figure~\ref{fig:networkLogistic9bits} presents the functional graph of Logistic map in a 9-bit floating-precision domain.

\begin{figure*}[!htb]
\centering
\begin{minipage}{\DoubleThreeImW}
\centering
\includegraphics[width=\DoubleThreeImW]{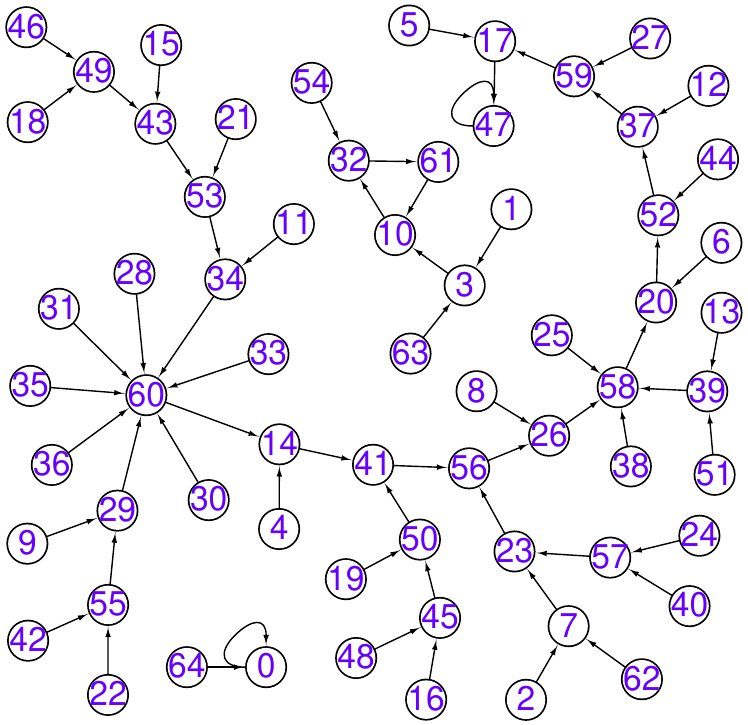}
a)
\end{minipage}\hspace{\figsep}
\begin{minipage}{\DoubleThreeImW}
\centering
\includegraphics[width=\DoubleThreeImW]{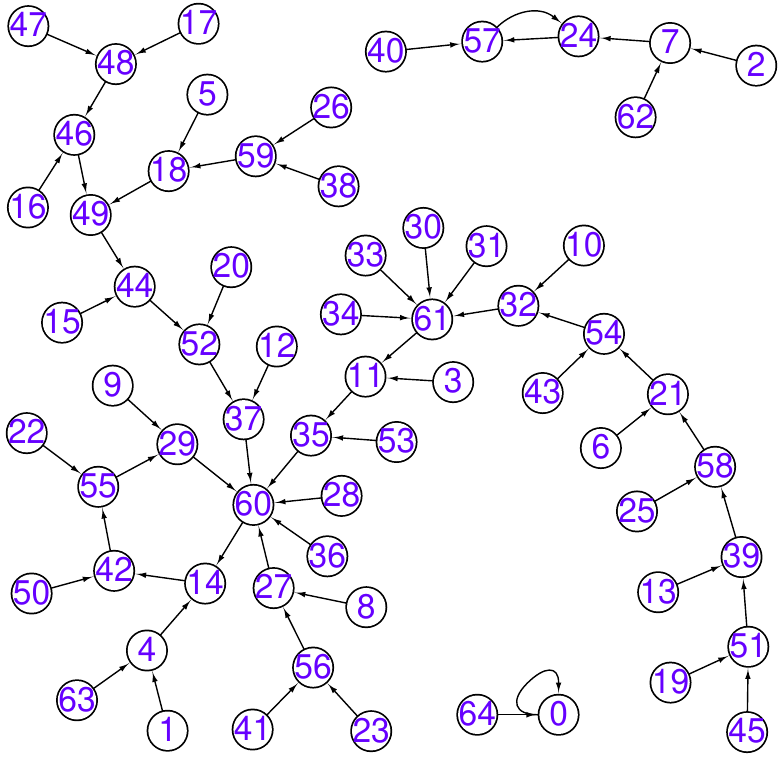}
b)
\end{minipage}\hspace{\figsep}
\begin{minipage}{\DoubleThreeImW}
\centering
\includegraphics[width=\DoubleThreeImW]{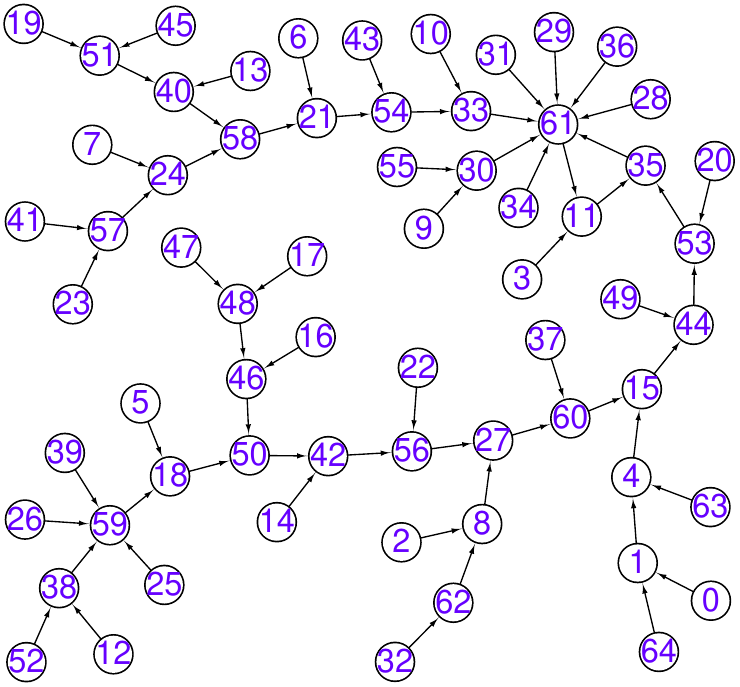}
c)
\end{minipage}
\caption{The functional graph of Logistic map with $\mu=61/2^4$ under 6-bit fixed-point precision for different quantization strategies:
a) floor; b) round; c) ceil, where the number $i$ in each node denotes value $i/2^6$.}
\label{fig:networkLogistic6bitsfloorroundceil}
\end{figure*}

\vspace{1cm}

\begin{figure*}[!htb]
\centering
\begin{minipage}{\DoubleThreeImW}
\centering
\includegraphics[width=\DoubleThreeImW]{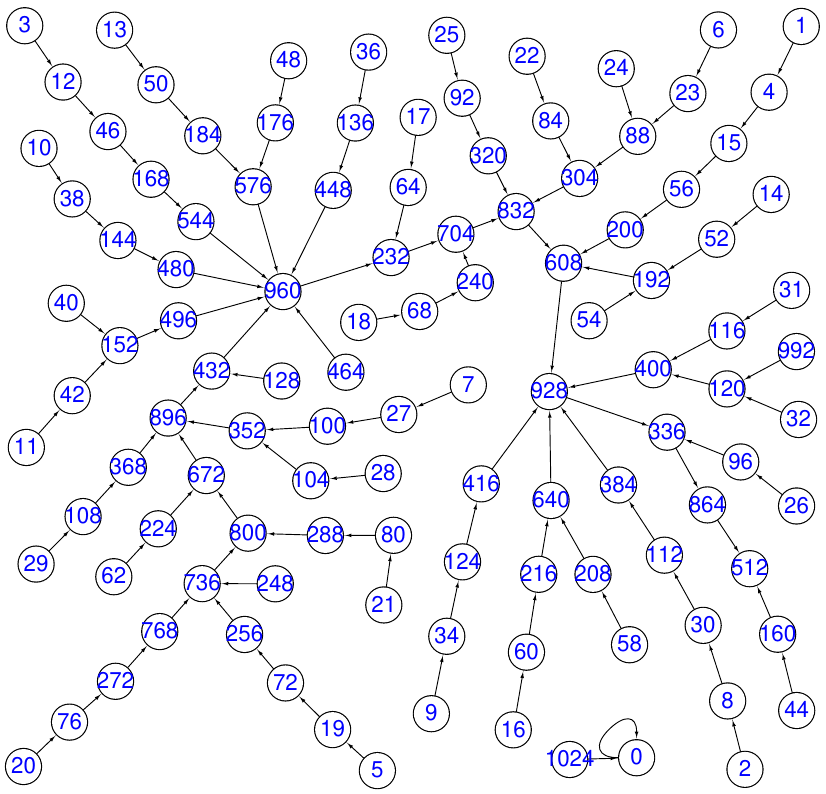}
\end{minipage}\hspace{\figsep}
\begin{minipage}{\DoubleThreeImW}
\centering
\includegraphics[width=\DoubleThreeImW]{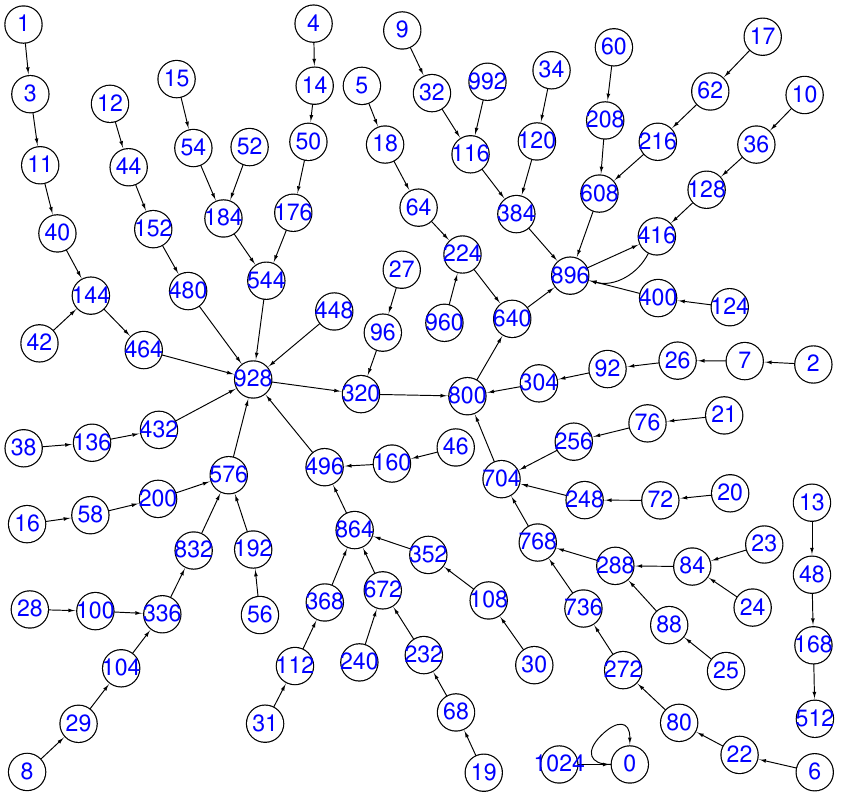}
\end{minipage}\hspace{\figsep}
\begin{minipage}{\DoubleThreeImW}
\centering
\includegraphics[width=\DoubleThreeImW]{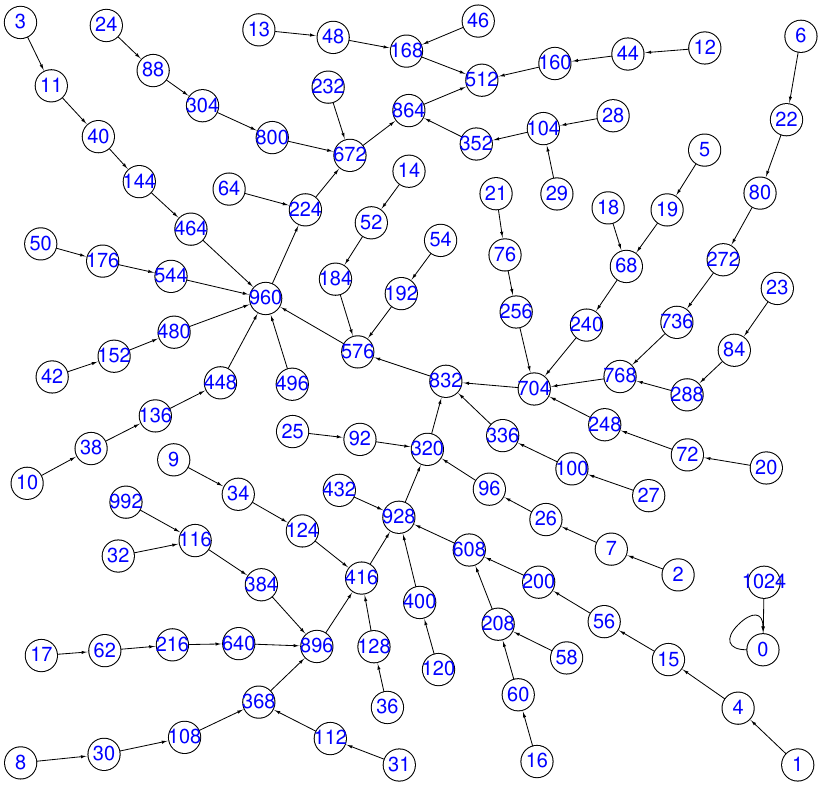}
\end{minipage}
\caption{The functional graph of Logistic map with $\mu=123/2^5$ under 9-bit floating-point precision,
where significant digits (the significand) and exponent both occupy 4 bits and the number $i$ in each node denotes value $i/2^{10}$.}
\label{fig:networkLogistic9bits}
\end{figure*}

\item Low efficiency of the method generating PRNS

In the field of image encryption, many schemes use
\begin{equation}
f_n(x)=f\left(10^m \cdot x\right) \bmod D
\label{multiply}
\end{equation}
to convert a floating-point number into an $n$-bit integer number, where
$f(x)$ is a quantization function, e.g. ceil, and floor.
In computer, complexity of multiplication of two $s$-bit binary numbers is
$O(s^\alpha)$, where $\alpha \in (1.35, 2]$ depending on the specific multiplication algorithm, e.g.,
Booth's algorithm and Karatsuba algorithm.
As
\[
10^m = \left( 2\cdot (2^2+1) \right)^m=\sum_{i=0}^m {m \choose i} \cdot 2^{3m-2i},
\]
the number of ``1" in the binary representation of $10^m$, $n_0$, is largely proportional to $m$.
In a machine adopting the ``shift and add" algorithm, the computational complexity is proportional to $n_0$.
Figure~\ref{fig:m_vs_n0} depicts how the binary length of $10^m$,  $\lceil \log_2(10^m) \rceil = \lceil m \log_2(10) \rceil$, and $n_0$ changes with respect to $m$ when $m\in [1, 50]$.
As for Eq.~(\ref{eq:conversion}) and Eq.~(\ref{eq:multiply14}), one has $m=14$, $n_0=17$, and only the eight and two least significant bits are adopted, respectively. So, computations on most computed bits are wasted.

\begin{figure}[!htb]
  \centering
  \begin{minipage}{\BigOneImW}
  \centering
  \includegraphics[width=\BigOneImW]{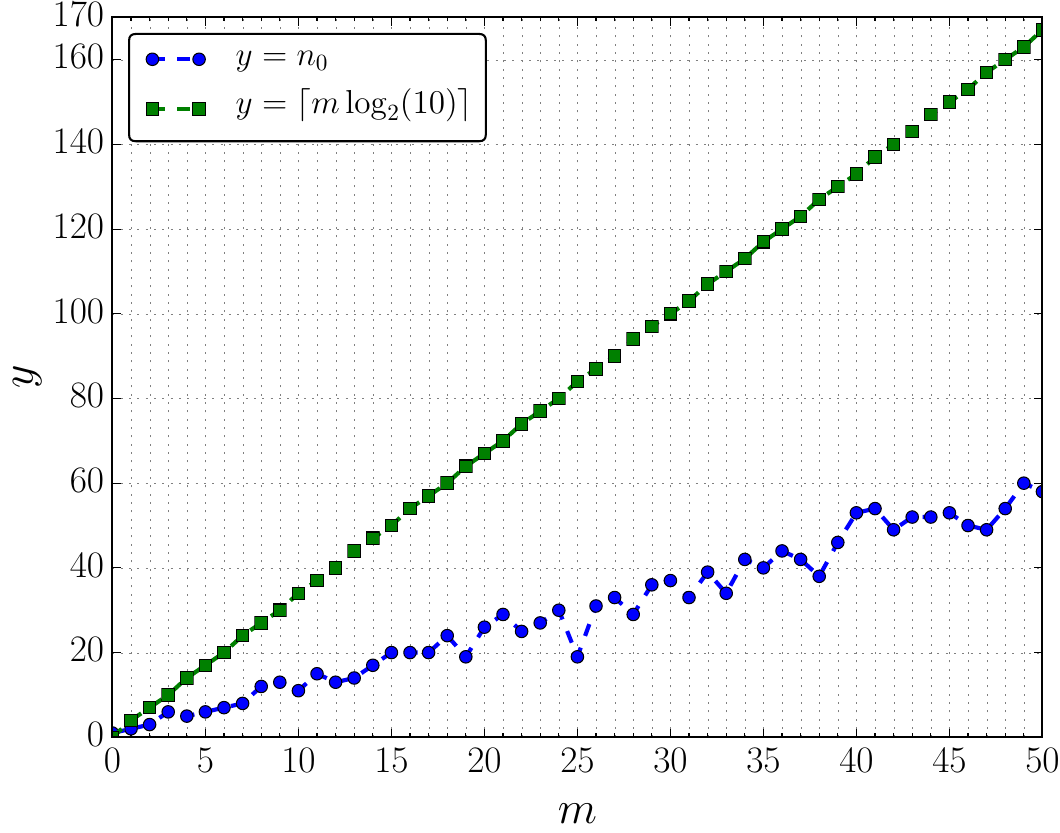}
  \end{minipage}
  \caption{The bit length of $10^m$ and the number of ``1" in its binary presentation.}
  \label{fig:m_vs_n0}
  \end{figure}

\item Period behavior of the generalized Arnold map

\begin{figure*}[!htb]
	\centering
	\begin{minipage}{2\BigOneImW}
		\centering
		\includegraphics[width=2\BigOneImW]{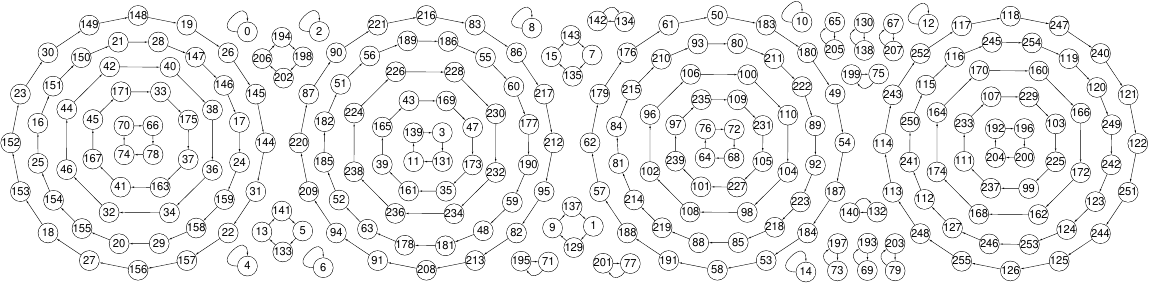}
		a)
	\end{minipage}
	\begin{minipage}{2\BigOneImW}
		\centering
		\includegraphics[width=2\BigOneImW]{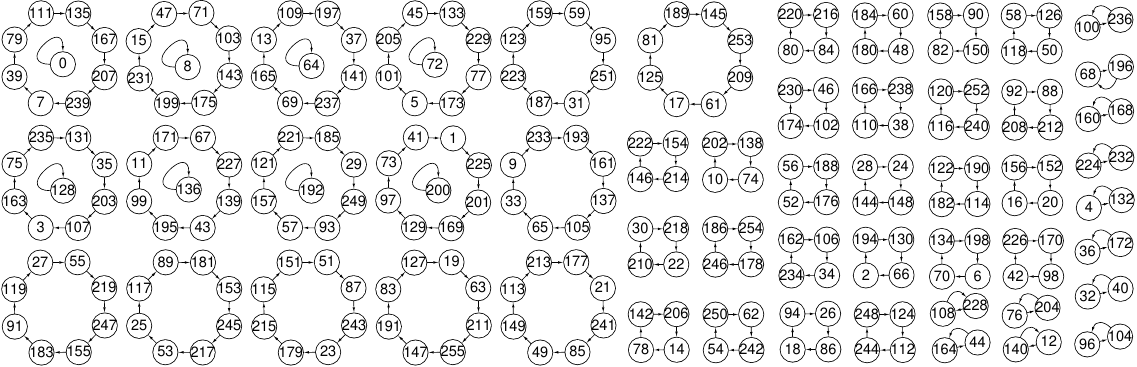}
		b)
	\end{minipage}
	\caption{Functional graphs of generalized Arnold maps in $\mathbb{Z}_{2^4}$: a) $a'=7$, $b'=8$; b) $a'=12$, $b'=14$.}
	\label{fig:SMNcat}
\end{figure*}

In $e$-bit fixed-precision arithmetical domain, Eq.~(\ref{eq:Arnold}) is equivalent to
\begin{equation}
\begin{pmatrix}
  x'_{n+1} \\
  y'_{n+1}
\end{pmatrix}
=
\begin{pmatrix}
    1 & a'     \\
    b' & 1+a'\cdot b'
\end{pmatrix}\cdot
\begin{pmatrix}
  x'_{n} \\
  y'_{n}
\end{pmatrix}
\bmod 2^e,
\label{eq:Arnold2}
\end{equation}
where $a'=a\bmod 2^e$, $b'=b\bmod 2^e$, $x'_n=\lfloor x_n\cdot 2^e \rfloor$, and $y'_n=\lfloor y_n\cdot 2^e \rfloor$.
To visualize the real structure of the generalized Arnold map, we depict its functional graph with two sets of parameters in Fig.~\ref{fig:SMNcat},
where the number in each node denotes
\[
z_n=x'_n+ (y'_n\cdot 2^e),
\]
and $e=4$. The whole graph shown in Fig.~\ref{fig:SMNcat}a) is composed of 8 connected components (CC) of period 16, 8 CC of period 8, 8 CC of period 4, 11 CC of period 2, and 8 self-connected nodes. So, the support of \cite[Fig.~2]{ye2016image} on random behavior of the generalized Arnold map is groundless. Actually, it only plots a short orbit (path) in a connected component obtained in the digital computer.
In addition, although the maximal Lyapunov exponent of the generalized Arnold map is positive, it is a metric measuring the overall dynamics of a system, which may fail to demonstrate complex dynamics (randomness) of the system in a local domain.

\item Incapability of the test metrics adopted by IEAE

\begin{itemize}[leftmargin=*]

\item \textbf{Key Space and Sensitivity}: The size of the key space of IEAE is bounded by the number of available of ECG signals, which incur
complex burdens of secure storage and transmission of the sensitive information and may violate the availability principle of security \cite{yap2016effective:VCIR}. Just like the dynamics of any chaotic system are degenerated in digital world \cite{WangQX:HDDCS:TCAS2016}, the sensitivity of digitized ECG signals with respect to a sampled person and time also worsens to some extents. Furthermore, due to the addition and division in Eq.~(\ref{lyapunov}), totally different ECG signals may own the same largest Lyapunov exponent.
So, the statement in \cite{ye2016image}, ``ECG is like a one-time keypad--different people will have different ECGs, so the keys will not be
used twice", is questionable.

\item \textbf{Histograms}: As shown in \cite{Cqli:Scramble:IM17}, an attacker cannot efficiently obtain some meaningful information
from the uniform histogram of pixels, but can learn important statistic information about the plain-image from the histogram of bits.
So, changing the objects of histogram may make the statement in \cite[Fig. 6]{ye2016image} become invalid.

\item \textbf{Differential Attack}: Differential attack is to find information about the secret key of an encryption scheme by studying how differences between plaintexts can affect the resultant difference between the corresponding ciphertexts, which is unrelated to the index $\mathit{UACI}$ measuring how plaintext influences the corresponding ciphertext.

\item \textbf{Correlation Coefficients}: Weak correlation among adjacent pixels is only a necessary (not sufficient) condition for an invisible cipher-image, which is only related with the capability resisting statistical analysis from a single cipher-image. In fact, position permutation is a sufficient way to reduce correlation coefficients among neighbouring elements in a plain-image \cite{LiShujun:Rules:IJBC2006}.

\item \textbf{Efficiency}: In \cite{ye2016image}, it is claimed that ``implementing encryption using the proposed algorithm is fast". In fact, the fast running speed of IEAE is built on the simple linear encryption function by sacrificing security. Note that the computational load spent on direct encryption of a plain-image is proportional to the number of plain-bytes and is unrelated with the block size. Even worse, a substantial part of the limited computational load of IEAE is wasted in the processes generating pseudorandom binary sequences. Among them, the computation of the largest Lyapunov exponent involves very complex operations (see Sec.~\ref{scheme} or \cite{wolf1985determining}) and the final computational complexity dependents on the specific used algorithmic. Furthermore, Wolf's method is also very dependant on the length of the time series and in parameters selection (i.e., the embedding dimension and the time delay). Such a dependence could incur differences between similar configurations in different computers or setups, which may lead to slightly different initial conditions of Logistic map. For example, the same configuration for computing
    logarithm in Eq.~(\ref{lyapunov}) has to be adopted by both secure communication sides. All these dependency problems erode the practicality of IEAE.
    To obtain a satisfying balance between efficiency, usability, and security, selecting some important data in the compressing domain of a plain-image for encryption is a practical approach.
\end{itemize}
\end{itemize}

\section{Conclusion}

This paper analyzed security of an image encryption algorithm based on autoblocking and electrocardiography, and showed that the algorithm was very weak against the known-plaintext attack. Security defects in the analyzed algorithm were summarized to inform
designers of image encryption schemes about common pitfalls and help them improve security levels protecting image data in the current cyberspace.

\section*{Acknowledgements}

This work was supported by the National Natural Science Foundation of China (no.~61772447, 61532020).

\bibliographystyle{IEEEtran}
\bibliography{Autoblocking}


\end{document}